\documentclass{trbunofficial}

\usepackage{xcolor}
\usepackage{xltabular}
\usepackage{array}
\usepackage{tabularx}
\usepackage{tikz}
\usetikzlibrary{arrows.meta, positioning, fit, backgrounds, calc}
\newcolumntype{Y}{>{\raggedright\arraybackslash}X}
  \makeatletter
  \def\LT@makecaption#1#2#3{%
    \LT@mcol\LT@cols l{\hbox to\z@{\parbox[t]\LTcapwidth{%
      \@makecaption{#2}{#3}\vskip\baselineskip}\hss}}}
  \makeatother

\makeatletter
\renewpagestyle{main}{%
  \sethead{\@AuthorHeaders}{}{}%
  \setfoot{}{\thepage}{}%
}
\patchcmd{\maketitle}{\thispagestyle{empty}}{\thispagestyle{main}}{}{%
  \ClassWarning{trbunofficial}{Could not number the title page}}
\makeatother

\begin{document}


\title{From Roadside Sensing to Risk-Informed Safety Control: Surrogate Safety Measures as a Cyber-Physical Interface}

\TRBauthor{Salsabil Salah}{PhD Student\\
Department of Civil Engineering\\
The City College of New York}{salsabil.salah54@stu-mail.ccny.cuny.edu}[New York, NY 10031]

\TRBauthor*{Yiqiao Li}{Assistant Professor\\
Department of Civil Engineering\\
The City College of New York\\
Computer Science Program\\
The City University of New York}{yli4@ccny.cuny.edu}[New York, NY 10031]

\TRBauthor{Bo Wang}{Assistant Professor\\
Department of Mechanical Engineering\\
The City College of New York}{bwang1@ccny.cuny.edu}[New York, NY 10031]

\AuthorHeaders{Salah, Li, and Wang}

\maketitle

\section{Abstract}

\noindent\textbf{Objectives:}~Proactive intersection safety analysis depends on trajectory data whose reliability varies by sensing modality, yet infrastructure-perception and surrogate safety measure (SSM) research have only recently begun to converge. This paper synthesizes both literatures alongside control theory to clarify how sensing choices, safety metrics, and predictive control connect end-to-end, and where infrastructure-derived risk signals could inform autonomous vehicle decision-making.
\hfill\break%
\noindent\textbf{Methods:}~{We searched journal articles, conference proceedings, preprints, and technical reports, primarily from 2015--2026, supplemented by citation chasing and revision-stage additions. Searches covered light detection and ranging (LiDAR), cooperative perception, time-to-collision (TTC), post-encroachment time (PET), extreme value theory, model predictive control (MPC), and signal control. Screening considered trajectory-based safety and control relevance, excluding detection benchmarks without safety linkage. This structured narrative synthesis retained 56 sources.}
\hfill\break%
\noindent\textbf{Findings:}~No single SSM is universally valid: one-dimensional TTC and modified TTC suit longitudinal conflicts, two-dimensional TTC extends time-based measurement to crossing geometries, and observed PET captures crossing and vulnerable road user (VRU) conflicts but only after the event. LiDAR consistently outperforms camera-only sensing for continuous three-dimensional trajectory extraction, illumination robustness, and privacy, though its range degrades in heavy rain, fog, and snow. Its output becomes control-ready only when paired with filtering—Kalman, Savitzky-Golay—that stabilizes noisy detections.
\hfill\break%
\noindent\textbf{Novelty:}~Recent work has linked {trajectories collected by autonomous vehicles (AVs)}, two-dimensional SSMs, and short-term crash risk. This review complements it from the infrastructure side: it examines how roadside sensing affects that chain's reliability, classifies SSMs by conflict geometry and temporal availability, and maps risk components onto control objectives and constraints for infrastructure and connected and automated vehicle (CAV) applications.
\hfill\break%
\noindent\textbf{Practical Applications:}~Findings can help agencies match sensing and metrics to specific conflict types, compose filtered measures into site-appropriate risk states, and judge how infrastructure-derived risk signals might feed signal systems, traffic management, and CAV planning stacks.

\newpage

\section{1. Introduction}\label{sec:intro}

Crash records remain the traditional basis for transportation safety evaluation, but they are reactive by definition: a crash must occur, be reported, and be geocoded before it can inform intervention. This lag is especially costly at intersections, where risky operational patterns such as misjudged gaps, occluded crossings, and aggressive right-turns-on-red can persist for years before producing enough crashes to be statistically visible. Proactive safety analysis, built on surrogate indicators extracted from everyday traffic interactions rather than collisions themselves, offers a faster and more diagnostic alternative.

Roadside sensing has become the primary enabler of this shift. Infrastructure-mounted cameras and, increasingly, roadside light detection and ranging (LiDAR) units provide persistent, site-level observation that can be converted into vehicle and pedestrian trajectories, and from there into conflicts and near-miss events, as discussed by \citet{sun2022object}. This capability has driven parallel growth in two research communities that have, until recently, developed largely independently of one another. The infrastructure-perception community has focused on detection accuracy, multi-object tracking, and cooperative-perception benchmarks, producing an expanding ecosystem of datasets and model architectures. The surrogate-safety community, meanwhile, has focused on defining and validating surrogate safety measures (SSMs) such as time-to-collision (TTC) and post-encroachment time (PET). \citet{arun2021systematic} note that this literature has largely treated the underlying trajectory data as a given input rather than as a variable with its own error structure.

\citet{steinmassl2025framework} discuss how this separation creates a practical problem for agencies attempting to deploy roadside sensing for safety decision-making: a perception system can perform well on conventional detection benchmarks and still generate trajectories too unstable---through occlusion, identity switching, or clock drift---to support valid conflict inference. Sensor choice, in other words, is not a preliminary engineering decision that can be separated from the safety analysis that follows it; it is part of the safety model itself.

A second, less-examined gap concerns what happens after a near-miss or conflict is identified. Most surrogate-safety work treats SSMs as offline evaluative outputs: metrics computed after the fact to rank sites, compare interventions, or feed crash-frequency models. A smaller but growing body of control-theoretic work, such as \citet{mihaly2021performance}, treats these same metrics as real-time signals: inputs to state estimators, constraints on trajectory optimization, and triggers for active intervention such as model predictive control (MPC) at intersections. This control-theoretic framing is directly relevant to {autonomous vehicles (AVs)}, which must translate ambient risk into actionable steering and braking decisions in real time. Yet it is rarely discussed alongside the roadside sensing and SSM literatures that would supply such systems with their input data. 

A small number of studies have begun to bridge these communities. \citet{li20241beyond} generalized SSMs from {one-dimensional (1D)} to multi-dimensional, higher-fidelity kinematics, and \citet{anis2025realtime, anis2026learning} built on such indicators to estimate short-term crash risk from AV-collected trajectories using hierarchical Bayesian extreme value models, including vehicle--infrastructure near misses along urban corridors. These studies establish that the sensing-to-risk chain can be quantified. Three questions remain less developed: how infrastructure-based, rather than vehicle-mounted, sensing affects the reliability of that chain; which SSMs become available early enough to support anticipatory rather than monitoring-level intervention; and how a risk representation should be structured so that controllers operating on different timescales and actuators can consume it.

This paper addresses these questions from the infrastructure side. We make three contributions. First, we compare roadside camera- and LiDAR-based sensing specifically for their suitability to safety-relevant trajectory extraction, rather than for detection accuracy alone. Second, we synthesize the SSM literature around a framework that matches metrics to conflict geometry and to temporal availability, distinguishing measures that can support anticipatory intervention from those observable only after an event concludes. Third, we extend the discussion from risk representation toward control, examining how filtered, scenario-matched surrogate measures and {crash-risk estimates based on extreme value theory (EVT)} can be composed into an actionable risk state, and how that representation can inform objectives and constraints for predictive and operational control, including connected and automated vehicle (CAV) decision-making.

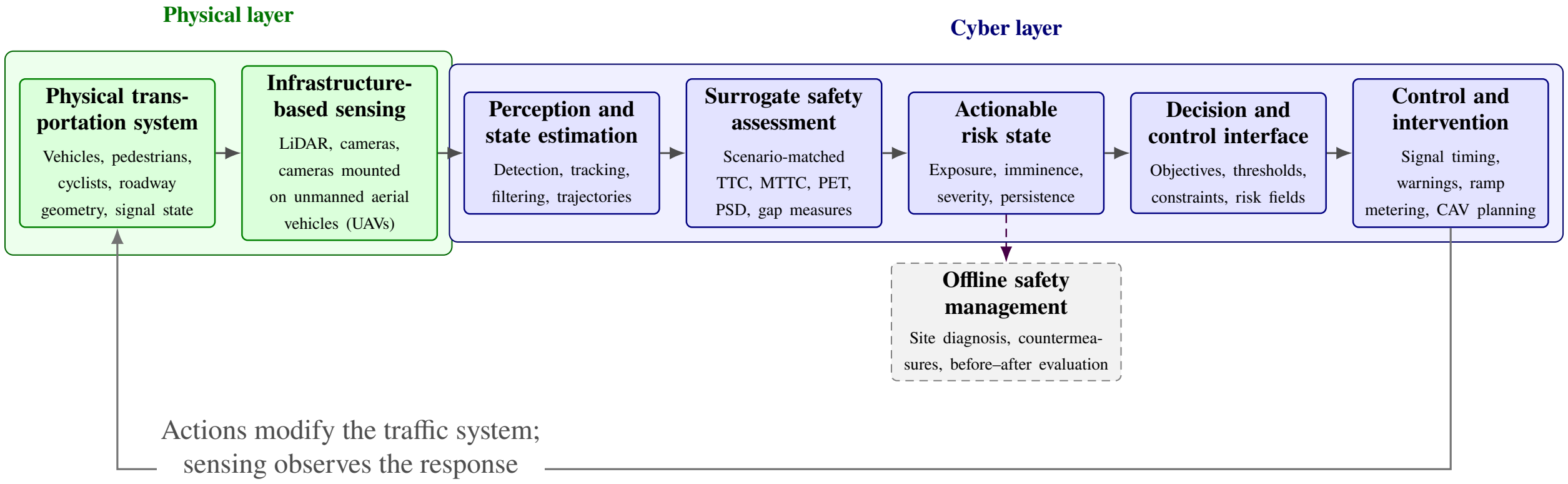
\begin{figure}[htbp]
\centering
\resizebox{\linewidth}{!}{%
\begin{tikzpicture}[
  >=Latex,
  font=\scriptsize,
  arr/.style={-{Latex[length=2mm,width=1.6mm]}, line width=0.65pt, draw=black!70},
  arrdash/.style={-{Latex[length=2mm,width=1.5mm]}, line width=0.55pt, draw=violet!55!black, densely dashed},
  arrfb/.style={-{Latex[length=2mm,width=1.6mm]}, line width=0.7pt, draw=black!55},
  phys/.style={
    rectangle, rounded corners=2pt, align=center, inner sep=3pt,
    minimum height=1.45cm, text width=0.88in,
    draw=green!50!black, fill=green!14, line width=0.55pt},
  cyber/.style={
    rectangle, rounded corners=2pt, align=center, inner sep=3pt,
    minimum height=1.45cm, text width=0.88in,
    draw=blue!50!black, fill=blue!11, line width=0.55pt},
  offline/.style={
    rectangle, rounded corners=2pt, align=center, inner sep=3pt,
    minimum height=1.05cm, text width=1.05in,
    draw=black!50, fill=black!5, line width=0.45pt, densely dashed},
]
\node[phys] (s1) {%
  \textbf{Physical transportation system}\\[2pt]
  \tiny Vehicles, pedestrians, cyclists, roadway geometry, signal state};
\node[phys, right=3.2mm of s1] (s2) {%
  \textbf{Infrastructure-based sensing}\\[2pt]
  \tiny LiDAR, cameras, cameras mounted on unmanned aerial vehicles (UAVs)};
\node[cyber, right=3.2mm of s2] (s3) {%
  \textbf{Perception and state estimation}\\[2pt]
  \tiny Detection, tracking, filtering, trajectories};
\node[cyber, right=3.2mm of s3] (s4) {%
  \textbf{Surrogate safety assessment}\\[2pt]
  \tiny Scenario-matched TTC, MTTC, PET, PSD, gap measures};
\node[cyber, right=3.2mm of s4] (s5) {%
  \textbf{Actionable risk state}\\[2pt]
  \tiny Exposure, imminence, severity, persistence};
\node[cyber, right=3.2mm of s5] (s6) {%
  \textbf{Decision and control interface}\\[2pt]
  \tiny Objectives, thresholds, constraints, risk fields};
\node[cyber, right=3.2mm of s6] (s7) {%
  \textbf{Control and intervention}\\[2pt]
  \tiny Signal timing, warnings, ramp metering, CAV planning};

\begin{scope}[on background layer]
  \node[fit=(s1)(s2), inner sep=5pt, rounded corners=3pt,
    fill=green!7, draw=green!40!black, line width=0.45pt] (physband) {};
  \node[fit=(s3)(s7), inner sep=5pt, rounded corners=3pt,
    fill=blue!6, draw=blue!40!black, line width=0.45pt] (cyberband) {};
\end{scope}

\node[anchor=south, font=\scriptsize\bfseries, text=green!45!black]
  at ([yshift=1.5mm]physband.north) {Physical layer};
\node[anchor=south, font=\scriptsize\bfseries, text=blue!45!black]
  at ([yshift=1.5mm]cyberband.north) {Cyber layer};

\draw[arr] (s1) -- (s2);
\draw[arr] (s2) -- (s3);
\draw[arr] (s3) -- (s4);
\draw[arr] (s4) -- (s5);
\draw[arr] (s5) -- (s6);
\draw[arr] (s6) -- (s7);

\node[offline, below=6mm of s5] (off) {%
  \textbf{Offline safety management}\\[1pt]
  \tiny Site diagnosis, countermeasures, before--after evaluation};
\draw[arrdash] (s5.south) -- (off.north);

\coordinate (fbY) at ([yshift=-11mm]off.south);
\draw[arrfb] (s7.south) -- (s7.south |- fbY) -- (s1.south |- fbY) -- (s1.south);
\coordinate (fbLbl) at ($(s1.south)!0.35!(s4.south)$);
\node[font=\footnotesize, align=center, text=black!70, inner sep=1.5pt,
  fill=white, draw=none]
  at ([yshift=2.5mm]fbLbl |- fbY)
  {Actions modify the traffic system;\\sensing observes the response};

\end{tikzpicture}%
}
\caption{Pipeline from roadside sensing (LiDAR, camera, UAV) through trajectory extraction, surrogate safety measures, and near-miss analytics, to control-theoretic action for autonomous vehicles. Green shading denotes the physical layer; blue shading denotes the cyber layer. Dashed branch: offline safety management fed by the actionable risk state. Return path: control actions alter the physical system, which sensing observes on the next cycle. {MTTC: modified time-to-collision; PSD: proportion of stopping distance.}}\label{fig:pipeline}
\end{figure}

The remainder of the paper proceeds along this arc. The {\customref{Review Methodology section}{sec:methodology}} describes how sources were identified and selected. The \customref{Sensing Technologies section}{sec:sensing} compares sensing modalities for safety-relevant data collection. The \customref{Detection, Tracking, and Trajectory Computation section}{sec:detection} discusses detection, tracking, and the computation of stable trajectories from raw sensor data. The \customref{Surrogate Safety Measures section}{sec:ssm} introduces the interaction--conflict--near-miss hierarchy and reviews SSMs and their scenario-specific validity, including how filtered, scenario-matched measures are aggregated into an actionable risk state. The \customref{Risk-State-Informed Safety Control section}{sec:control} develops that control thread, connecting kinematic estimation and risk-based constraints to illustrative control mechanisms and application domains for CAVs. The \customref{Future Directions section}{sec:future} outlines directions for future research before the \customref{Conclusion}{sec:conclusion}. Figure~\ref{fig:pipeline} summarizes this pipeline.

\section{2. Review Methodology}
\label{sec:methodology}

This paper is a structured narrative synthesis rather than a systematic review. The three literatures it connects (roadside perception, surrogate safety measurement, and safety-oriented control) use largely non-overlapping vocabularies. A single Boolean query would therefore either miss relevant work or return an unmanageable volume of detection benchmarks. We instead searched each pairing of domains separately and supplemented the results with citation chasing.

\textit{Sources and period.} We searched Scopus, Web of Science, TRID, IEEE Xplore, Google Scholar 
and the arXiv and SSRN preprint servers for journal articles, proceedings, preprints, and technical reports published between January 2015 and August 2026. 
The final search was run on August 1, 2026. {Additional references on safety-critical control were incorporated during manuscript revision, including work available after this search date.} 
Foundational earlier work \citep[e.g.,][]{hayward1972nearmiss, horst1994time, archer2005indicators, laureshyn2010evaluation} was added through backward citation chasing.

\textit{Search terms.} Terms were grouped into three blocks:
\begin{itemize}
    \item[(A)] \textit{Sensing:} ``roadside LiDAR,'' ``infrastructure-based sensing,'' ``cooperative perception,'' ``UAV trajectory,'' ``autonomous vehicle sensor data'';
    \item[(B)] \textit{Safety measures:} ``surrogate safety measure,'' ``time-to-collision,'' ``two-dimensional TTC,'' ``post-encroachment time,'' ``traffic conflict,'' ``near miss,'' ``extreme value theory,'' ``real-time crash risk'';
    \item[(C)] \textit{Control:} ``model predictive control,'' ``adaptive signal control,'' ``collision avoidance constraint,'' ``risk field.''
\end{itemize}
\noindent Blocks were combined pairwise ($A \wedge B$, $B \wedge C$, $A \wedge C$).

\textit{Eligibility.} Studies were included if they addressed intersection or urban-corridor safety and did at least one of the following: extracted road-user trajectories for safety analysis, defined or validated an SSM or conflict-based risk estimate, or used a safety measure as a control objective or constraint. We excluded detection or tracking benchmarks with no safety linkage, crash-frequency studies without surrogate measures, and non-English sources.

\textit{Screening and synthesis.} The authors screened titles and abstracts, and reviewed candidate full texts; disagreements were resolved by discussion. 
In total, {56} sources were retained for the synthesis.

\textit{Limitations.} Selection followed no pre-registered protocol, and no formal quality appraisal was applied, so some relevant work may be under-represented.

\section{3. Sensing Technologies for Roadside Safety Data Collection}\label{sec:sensing}

This section evaluates the sensing technologies available for extracting safety-relevant vehicle and pedestrian trajectories at intersections, since the fidelity of that upstream data bounds what any downstream SSM or control system can reliably compute. Cameras, roadside LiDAR, and unmanned aerial vehicle (UAV) platforms are compared not on raw detection accuracy but on their suitability for continuous, safety-grade trajectory extraction under real-world coverage, weather, computational, and privacy constraints. The section closes with a comparative assessment and a discussion of multimodal fusion, setting up the detection and tracking pipelines examined in the \customref{Detection, Tracking, and Trajectory Computation section}{sec:detection}.

\subsection{Camera-Based Sensing}\label{sec:camera}

Roadside cameras have historically served as the default sensor for intersection monitoring, valued for low hardware cost and strong two-dimensional object recognition. However, extracting reliable trajectories from camera video requires converting a {two-dimensional (2D)} pixel plane into {three-dimensional (3D)} world coordinates, a step that introduces measurement error before a single SSM is even computed. \citet{guan2023evaluation} compare trajectory data extracted from roadside LiDAR against trajectory data extracted from computer vision systems and find that estimating depth from two-dimensional camera pixels demands multiple stereo cameras and complex triangulation. \citet{sun2022object} similarly note that camera-based systems are highly vulnerable to darkness and abrupt changes in illumination. Achieving the continuous, wide-area depth coverage that safety analytics requires {may demand} overlapping camera arrays, an approach which grows costly and may be difficult to maintain as detection range increases.

\subsection{Roadside LiDAR Sensing}\label{sec:lidar}

Roadside LiDAR addresses these limitations by measuring physical distance directly rather than inferring it from pixel geometry. \citet{lin2023vehicle} describe how infrastructure-mounted LiDAR units scan their environment through a full 360-degree field of view in real time, producing wide-area coverage and detailed depth information suited to high-resolution micro-traffic data extraction. \citet{zhang2022traffic} quantify this capability, reporting that a standard roadside unit such as the Velodyne VLP-16 provides a 360-degree horizontal field of view with a continuous scanning radius of up to 100 meters. Because LiDAR captures physical geometry directly as point clouds, \citet{guan2023evaluation} find that it avoids the multi-camera triangulation overhead associated with vision-based systems while also extending detection range and remaining largely unaffected by lighting conditions.

LiDAR also offers a privacy advantage that is increasingly relevant to agency deployment decisions. \citet{igene2024integrating} contrast video-based sensing, which captures detailed imagery capable of revealing personally identifiable information such as faces and license plates, with LiDAR, which represents road users purely as anonymous spatial point clouds. This distinction can ease the regulatory and public-acceptance burden of deploying persistent roadside monitoring infrastructure.

These advantages are not unconditional. Adverse weather remains a genuine constraint for LiDAR: rain, fog, and snow scatter laser pulses and generate spurious return points, degrading both effective range and reflection intensity. \citet{xu2018high} note that even with this constraint, roadside LiDAR still outperforms camera-only sensing in crowded, illumination-variable scenes, since pulsed-laser ranging allows point clouds to be clustered and classified into distinct road-user types using direct spatial coordinates rather than macro-level, lighting-sensitive image features.

\subsection{Emerging Aerial and Complementary Modalities}\label{sec:aerial}

Beyond fixed roadside sensors, UAVs have emerged as a flexible, bird's-eye-view alternative for trajectory-based safety analysis. \citet{espadaler2025accurate} deploy a swarm of drones equipped with 4K cameras to compute TTC and PET across two busy signalized intersections in Manchester, {United Kingdom}, using an anchor-free {YOLOv8 detector from the You Only Look Once (YOLO) family} paired with BoT-SORT tracking to extract stabilized trajectories. \citet{maji2025safety_a} apply similar drone-derived trajectory extraction across 23 heterogeneous roundabouts in India, computing not only TTC and PET but also {time-exposed TTC and time-integrated TTC} to characterize conflict risk in non-lane-based, highly mixed traffic. \citet{bhattarai2024drone} points toward a more integrated future, pairing UAV-mounted LiDAR and visual sensors with edge computing to reduce processing latency for real-time operational safety and emergency response.

This flexibility comes with constraints that limit UAVs as a substitute for persistent infrastructure. Flight endurance restricts most studies to short observation campaigns rather than continuous monitoring, and platform instability from wind can distort trajectory extraction unless explicitly corrected; \citet{shawky2023traffic} address this directly, applying a nonlinear optimization method to correct for drone instability before extracting conflict trajectories at a roundabout in New Cairo, Egypt. Airspace regulation further constrains routine deployment near active roadways. These constraints are why roadside LiDAR, suited to continuous, unattended monitoring, remains this paper's primary focus, with UAV-based sensing treated here as a complementary rather than substitute technology.

\subsection{Comparative Assessment}\label{sec:sensing-compare}

Table~\ref{tab:sensing-comparison} extends the earlier comparison to include UAV-based sensing alongside cameras and LiDAR, summarizing trade-offs across the dimensions most relevant to safety-oriented trajectory extraction. No single sensor dominates on every dimension: fixed cameras remain cheapest and offer intuitive two-dimensional classification; roadside LiDAR offers consistent geometric accuracy, weather-independent ranging outside of heavy precipitation, and privacy-preserving representation; and UAVs offer rapid, infrastructure-free deployment and a bird's-eye vantage well suited to complex or non-lane-based geometries, at the cost of limited flight endurance and exposure to wind-induced trajectory distortion. This pattern motivates the fusion approaches discussed below and directly shapes which sensor a deployment should prioritize for the conflict analysis discussed in the \customref{Surrogate Safety Measures section}{sec:ssm}.

\begin{table}[!ht]
\caption{Comparison of Camera-, LiDAR-, and UAV-Based Roadside Sensing for Safety-Relevant Trajectory Extraction}\label{tab:sensing-comparison}
\centering
\small
\begin{tabularx}{\linewidth}{@{}>{\raggedright\arraybackslash}p{0.17\linewidth}YYY@{}}
\hline
Dimension & Camera & LiDAR & UAV / Drone \\
\hline
Coverage / range & Requires overlapping arrays for wide coverage & Native 360\textdegree; up to $\sim$100~m radius per unit & Wide, flexible bird's-eye coverage; limited by flight time \\
Depth accuracy & Requires stereo triangulation; error-prone & Direct 3D distance measurement & Pixel-based; requires homographic correction \\
Weather robustness & Degrades with glare, shadow, low light & Resilient in normal conditions; degrades in heavy rain, fog, snow & Wind-sensitive; limited operation in adverse weather \\
Privacy & Captures faces, plates (personally identifiable information) & Anonymous point clouds only & Captures identifiable imagery, similar to camera \\
Computational cost & High (multi-camera triangulation) & Lower for trajectory extraction; higher for point-cloud processing & Moderate (video-based detection and tracking) \\
Hardware / deployment cost & Lower; permanent installation & Higher; permanent installation & Lower per-unit cost; non-persistent, requires piloting and regulatory compliance \\
\hline
\end{tabularx}
\end{table}

\subsection{Multimodal and Fusion Approaches}\label{sec:fusion}

A growing share of roadside perception systems do not treat these sensors as mutually exclusive choices. Bird's-eye-view fusion frameworks combine LiDAR geometry with camera semantics, and cooperative-perception designs extend this further by combining fixed roadside sensors with onboard vehicle sensors or neighboring infrastructure nodes. The aerial-ground pairing introduced in the \customref{Emerging Aerial and Complementary Modalities section}{sec:aerial} and the integration by \citet{bhattarai2024drone} of UAV-mounted LiDAR, visual sensors, and edge computing {suggest this fusion logic} is beginning to extend to mobile platforms as well, not just fixed installations.

Taken together, this section shows that sensing choice is not a neutral preliminary step but a decision that propagates directly into trajectory quality, and from there into every safety metric and control action built on top of it. Camera, LiDAR, and UAV sensing each carry distinct coverage, accuracy, weather, privacy, and cost trade-offs, with LiDAR generally best suited to the continuous, unattended, weather-resilient monitoring that proactive safety analysis requires, and fusion approaches increasingly used to offset any single modality's weaknesses. The next section turns from what these sensors capture to how that raw data is converted into the discrete, trackable trajectories that safety analysis depends on.

\section{4. Detection, Tracking, and Trajectory Computation}\label{sec:detection}

With a sensing modality selected, raw detections must still be converted into discrete, temporally continuous trajectories before any safety analysis is possible. This section reviews the detection architectures and tracking pipelines used to accomplish this for both LiDAR- and camera/UAV-based systems, and examines how filtering methods address the noise, occlusion, and identity-switching problems that most directly threaten trajectory quality. The goal throughout is to identify not just what makes a pipeline accurate on conventional benchmarks, but what makes it reliable enough to support the conflict inference introduced in the \customref{Surrogate Safety Measures section}{sec:ssm}.

\subsection{3D Detection Architectures for LiDAR Point Clouds}\label{sec:lidar-detect}

Converting a raw LiDAR point cloud into discrete actors is the first algorithmic step in the safety pipeline, and roadside detectors inherit much of their architecture from the broader 3D object detection literature. Point-based and voxel-based detectors including VoxelNet \citep{zhou2018voxelnet}, PointRCNN \citep{shi2019pointrcnn}, PointPillars \citep{lang2019pointpillars}, PV-RCNN \citep{shi2020pvrcnn}, CenterPoint \citep{yin2021center}, Voxel R-CNN \citep{deng2021voxel}, CenterFormer \citep{zhou2022centerformer}, and {Dynamic Sparse Voxel Transformer (DSVT)} \citep{wang2023dsvt} remain representative building blocks for extracting actors from sparse point clouds. Bird's-eye-view fusion frameworks such as BEVFusion \citep{liu2024bevfusion} extend this stack when cameras or cooperating sensors are present alongside LiDAR. \citet{kim2025lidar} illustrate how these architectures translate into applied safety pipelines: their framework pairs a Voxel R-CNN detector with a Contextual Intersection-over-Union tracking scheme that adjusts its association threshold dynamically based on vehicle length and relative speed, using a Pandar128 LiDAR mounted on a probe vehicle across South Korean highway and urban corridors. Their pipeline outputs two purpose-built indicators, a Hazardous Modified Time-to-Collision and a Searching-for-Safety-Space measure.

\subsection{Camera- and UAV-Based Detection and Tracking Pipelines}\label{sec:cam-detect}

Camera- and UAV-based studies rely on a parallel but architecturally distinct pipeline, typically pairing a two-dimensional object detector with a lightweight tracking-by-detection algorithm. \citet{espadaler2025accurate} pair an anchor-free YOLOv8 detector with BoT-SORT tracking, enhanced with Kalman filtering, to extract stabilized vehicle trajectories from swarm-collected drone footage. In another study, \citet{herjuna2026yolo} use a fine-tuned YOLO11 model combined with ByteTrack for pedestrian trajectory extraction, while \citet{mussah2024video} apply a modified YOLOv7 detector with ByteTrack on legacy closed-circuit television (CCTV) infrastructure. Because these pipelines inherit the two-dimensional depth limitation discussed in the \customref{Camera-Based Sensing section}{sec:camera}, several studies apply a bird's-eye-view homographic transformation prior to trajectory extraction; \citet{mussah2024video} and \citet{roy2025video} both use this correction to convert pixel-plane detections into spatially consistent coordinates before computing conflict metrics.

This parallel pipeline underscores a point raised earlier: sensing modality and processing pipeline are not independent choices. A camera- or UAV-based deployment inherits the depth-estimation burden identified earlier directly into its tracking stage, while a LiDAR-based deployment defers that burden to upstream sensor geometry instead.

\subsection{Multi-Object Tracking and Trajectory Quality in Roadside LiDAR}\label{sec:tracking}

Detection identifies where an actor is in a single frame; safety analysis requires knowing where that actor has been and where it is going. \citet{weng20203dmot} establish baseline evaluation metrics for 3D multi-object tracking and note that tracking frameworks are especially sensitive to identity switches, partial trajectories, and short-lived misses---precisely the failure modes that matter most for safety analysis, since they tend to cluster around conflict zones where two actors are actively negotiating priority.

A cluster of recent roadside LiDAR studies has responded to this sensitivity by treating trajectory quality itself as the evaluation target. \citet{chang2023using} use spatiotemporal stacks to achieve more precise vehicle tracking than frame-by-frame detection allows, while \citet{zhang2020vehicle} combine an Unscented Kalman Filter with a Joint Probabilistic Data Association Filter for centroid tracking, followed by a refinement step that projects 3D vehicle clusters into 2D images for pixel-level speed matching, independently validated against a vehicle instrumented with a global navigation satellite system (GNSS) receiver and an inertial measurement unit (IMU). \citet{zhang2025anchor} develop an anchor-free tracking method that identifies vehicles by center point rather than bounding box, pairing {Simple Online and Realtime Tracking (SORT)} with a Kalman filter and a spatial-index-based Hungarian association step to improve real-time trajectory continuity.

Occlusion remains the persistent failure mode across this literature. \citet{zhao2025optimizing} address point sparsity and occlusion directly in low-channel roadside LiDAR, proposing a joint clustering-and-classification method paired with an adaptive current statistical model that predicts vehicle position through full occlusion using a decaying factor, extending consistent trajectory extraction to a 117-meter range. \citet{igene2024integrating} integrate this filtering logic into a full deployment pipeline, combining density-based clustering, neural-network classification, and Global-Nearest-Neighbor tracking with a low-pass Savitzky-Golay filter to smooth velocity and acceleration profiles before computing the modified time-to-collision (MTTC), Stopping Distance Index, and Crash Severity Index that together form their Rear-End Conflict Index.

Together, these studies reflect a broader shift in the roadside perception literature: the pipeline must not only identify actors, but must produce trajectories with enough temporal continuity, synchronization, and geometric consistency to support conflict extraction. Kalman filtering, in one variant or another, appears as a {common component across these tracking algorithms}, a detail we revisit in the \customref{Risk-State-Informed Safety Control section}{sec:control}, where filtering functions not just as a tracking aid but as a precondition for using trajectory data as a real-time control signal.

\subsection{Implications for Downstream Safety Analysis}\label{sec:downstream}

Together, these detection and tracking choices determine whether a deployment's output is merely a stream of actor positions or a set of trajectories reliable enough to support conflict inference. A pipeline can score well on conventional detection and tracking benchmarks---mean average precision, multi-object tracking accuracy---and still be unsuitable for safety inference if trajectories fragment at the exact moment two actors negotiate priority, braking, or yielding. This distinction, between perception-valid and safety-valid output, is the foundation for the SSMs introduced next. It holds across both LiDAR- and camera-based pipelines alike: filtering is not an optional refinement but the mechanism that turns a merely accurate detector into a safety-valid one.

\section{5. Surrogate Safety Measures}\label{sec:ssm}

Having established how stable trajectories are produced, this section turns to how those trajectories are converted into quantitative safety indicators. It introduces the interaction--conflict--near-miss hierarchy that structures the surrogate-safety literature, then works through the specific metrics used within that hierarchy and the traffic scenarios each is best suited to, closing with a discussion of how these ordinarily offline metrics can instead function as real-time signals.

\subsection{The Interaction--Conflict--Near-Miss Hierarchy}\label{sec:hierarchy}

Once stable trajectories are available, the output of the pipeline described in the \customref{Detection, Tracking, and Trajectory Computation section}{sec:detection} offers a structured vocabulary for converting them into proactive risk indicators. \citet{hayward1972nearmiss} introduced the foundational concept of a scale of danger to distinguish routine traffic encounters from those warranting explicit safety attention, a distinction later formalized by \citet{archer2005indicators} and by \citet{laureshyn2010evaluation} into the three-tier hierarchy used throughout the modern surrogate-safety literature. Interactions are ordinary multi-actor encounters that occur continuously in normal traffic flow. Conflicts are the subset of interactions that approach a collision condition under some behavioral or kinematic model. Near misses are, in turn, the subset of conflicts judged severe enough to merit explicit safety review, typically based on a threshold applied to an SSM.

This hierarchy matters beyond definition: many datasets and sensing deployments can support interaction mining, since ordinary encounters are common and comparatively easy to log, but far fewer can support defensible conflict inference or credible near-miss labeling, both of which require the trajectory precision and continuity discussed in the \customref{Detection, Tracking, and Trajectory Computation section}{sec:detection}.

\subsection{Matching Safety Metrics to Traffic Scenarios}\label{sec:matching}

As the \customref{Sensing Technologies section}{sec:sensing} and \customref{Detection, Tracking, and Trajectory Computation section}{sec:detection} establish, evaluating proactive safety requires sensing and trajectory pipelines capable of continuous, unbroken data under challenging conditions. Possessing accurate trajectory data is only the first step, however; \citet{mahmud2017application} emphasize that the chosen metric must be carefully matched to the operational geometry and kinematic dynamics of the specific traffic encounter. \citet{johnsson2018search} illustrate this concretely: a metric designed for stable longitudinal car-following will fail to capture the lateral evasive maneuvers, stop-and-go acceleration, or complex pedestrian yielding behaviors common at urban intersections. Table~\ref{tab:ssm-mapping} outlines the operational scenarios for the SSMs most commonly used in LiDAR-based proactive safety pipelines.
The TTC, MTTC, and PSD formulations in Table~\ref{tab:ssm-mapping} assume motion along a single longitudinal axis. \citet{li20241beyond} showed that these one-dimensional measures are special cases of a general framework that predicts both road users' footprints under higher-order kinematics and defines time-to-collision as the earliest predicted overlap. They also found that 1D values can deviate substantially from higher-dimensional ones, by roughly 20\% even for critical TTC values below 1.5~s. Such two-dimensional formulations extend time-based metrics beyond longitudinal conflicts to crossing, turning, and lane-change geometries. Their main requirement is continuous heading and footprint estimates, which ties metric choice back to the sensing and tracking choices discussed in the preceding sections.

\begingroup
\linespread{0.93}\selectfont
\renewcommand{\arraystretch}{1.05}
\setlength{\tabcolsep}{4pt}
\setlength{\LTcapwidth}{\linewidth}
\begin{xltabular}{\linewidth}{@{}>{\footnotesize\raggedright\arraybackslash}p{0.17\linewidth}>{\footnotesize\hsize=1.15\hsize}Y>{\footnotesize\hsize=0.90\hsize}Y>{\footnotesize\hsize=0.95\hsize}Y@{}}
\caption{Mapping Widely Used Surrogate Safety Measures to Their Characteristics}\label{tab:ssm-mapping}\\
\hline
{Metric / temporal availability} & Measure & Ideal Traffic Scenario & Why LiDAR? \\
\hline
\endfirsthead
\caption[]{(continued)}\\
\hline
{Metric / temporal availability} & Measure & Ideal Traffic Scenario & Why LiDAR? \\
\hline
\endhead
\hline
\endfoot
\hline
\endlastfoot
TTC (conventional 1D time-to-collision)\par\smallskip{\textit{Anticipatory}} &
Evaluates longitudinal gap between a leading and following vehicle divided by velocity difference ($V_{\text{follow}} > V_{\text{lead}}$), assuming linear kinematics along a single longitudinal axis. A common threshold used is 1.5 seconds \citep{li20241beyond}. &
A good measure for rear-end conflicts in highway segments or straight urban corridors; applicable across automated and human-driven vehicles \citep{horst1994time}. &
LiDAR 3D point clouds have spatial positioning and depth, which can be used for this, but with cameras, {adverse weather may reduce depth accuracy} \citep{ozbek2021performance}. \\
MTTC (modified TTC)\par\smallskip{\textit{Anticipatory}} &
Extends TTC by accounting for relative acceleration and deceleration between leading and following vehicles \cite{ozbay2008derivation}. &
Applicable for dynamic conditions, stop-and-go traffic, complex link-based analyses, and scenarios involving larger commercial vehicles beyond standard passenger cars. &
Advanced measures can be accurately computed using data from LiDAR. \\
2D TTC (two-dimensional time-to-collision)\par\smallskip{\textit{Anticipatory}} &
Predicts the future footprints (dimensions and heading) of both road users under higher-order kinematics and returns the earliest time they overlap; conventional 1D TTC is a special case \citep{li20241beyond, behera2025improved}. &
Crossing, turning, lane-change, and curved-alignment conflicts where 1D TTC is undefined or biased; also vehicle--infrastructure conflicts with curbs and barriers \citep{anis2026learning}. &
Requires continuous position, heading, and footprint estimates, which LiDAR point clouds provide directly. \\
PET (post-encroachment time)\par\smallskip{\textit{Event-completion only}} &
Calculated as the time interval between the first road user leaving a conflict area and the second road user subsequently entering the same area. Acts as a margin of safety between two users with low sensitivity to trajectory errors. A generalized threshold of up to 6~s is considered \citep{singh2024conflict}, with typical vehicles occupying a 3.5~m $\times$ 3.5~m area while turning. &
Good for conflicts at intersections, crossings, and mixed traffic environments such as developing countries \citep{islam2023effect}; applicable to pedestrians, three-wheelers, and heavy commercial vehicles \citep{ranjan2026assessment}. &
\citet{mahmud2017application} note that traditional video-based approaches used to extract metrics like PET suffer from camera adjustment challenges, restricted coverage, and inaccuracies when transforming data from a 3D world into a 2D image plane. \\
IAPT (initially attempted post-encroachment time)\par\smallskip{\textit{Intermediate}} &
Calculated as the time lapse between the commencement of encroachment by a turning vehicle plus the expected time for the through vehicle to reach the point of collision, and the completion time of encroachment by the turning vehicle. &
Applies to a specific scenario defined by the Federal Highway Administration \cite{gettman2003surrogate}: turn-conflict events where an encroaching vehicle crosses in front of traffic that has the right-of-way. &
Same as PET. \\
PSD (proportion of stopping distance)\par\smallskip{\textit{Anticipatory}} &
Calculated as the distance to {the conflict point} divided by the minimum stopping distance. If PSD $>$ 1, the situation is safe; otherwise, there is not enough physical space to stop safely \citep{westhofen2023criticality}. &
Similar to TTC; applied in rear-end and car-following scenarios along a straight, longitudinal path where only braking can help avoid a collision. &
LiDAR provides continuous, unbroken physical distance measurements unaffected by camera glare or lighting transitions, preventing the depth-estimation errors that plague video-derived stopping-distance and gap calculations. \\
Gap time\par\smallskip{\textit{Anticipatory}} &
Calculated as the predicted separation from current speeds and {positions} of two vehicles. Values are determined using instantaneous speed and distance; functions for both cars and trucks. &
Ideal for left-turn scenarios to account for uncertainty in judging gaps between left-turning vehicles and opposing traffic, particularly useful for analyzing drivers with sight issues \citep{duong2010mechanistic}. &
Capturing precise vehicle entry and exit timestamps in intersection conflict zones requires continuous, high-resolution 3D trajectory tracking unaffected by lighting conditions and depth distortions. \\
\hline
\multicolumn{4}{@{}>{\footnotesize\raggedright\arraybackslash}p{\linewidth}@{}}{%
\textit{Note:} Temporal availability indicates when a measure's value becomes known.
\textit{Anticipatory}: computed from current positions and velocities while the interaction is still unfolding, so it can support warnings or evasive action within the same event.
\textit{Event-completion only}: defined only after the event concludes (for observed PET, once the second road user enters the conflict area), so it supports monitoring and later control actions but cannot give advance warning of that same event.
\textit{Intermediate}: known once the encroaching vehicle clears the conflict area, before the through vehicle arrives, but relies on a predicted arrival time.}\\
\end{xltabular}
\endgroup

\subsection{From Near Misses to Crash Risk: Extreme Value Approaches}
\label{sec:evt}

Thresholded SSMs identify near misses but do not, by themselves, estimate how likely a crash is. {EVT} addresses this by fitting the tail of an SSM distribution and extrapolating it to the collision boundary. \citet{maji2025safety_a} and \citet{roy2025video}, whose sensing pipelines are discussed above, apply EVT to drone- and CCTV-derived conflicts. \citet{anis2025realtime} combined high-fidelity 2D TTC computed from AV perception data with hierarchical Bayesian generalized extreme value models to estimate real-time crash risk across six arterial networks in San Francisco, Phoenix, and Los Angeles. \citet{anis2026learning} extended this approach to corridor-wide crash-occurrence risk, adding geometry-aware vehicle--infrastructure near misses and roadway covariates. Together, these studies operationalize a quantitative pipeline from high-resolution trajectories through 2D SSMs and near misses to short-term crash risk. Two features distinguish them from the focus of this review: their trajectories come from vehicle-mounted rather than fixed roadside sensors, and their output is a statistical crash-risk estimate rather than a control objective or constraint.

\subsection{From Surrogate Measures to an Actionable Risk State}\label{sec:riskstate}

Individual SSMs describe particular dimensions of an interaction but do not, by themselves, constitute a complete representation of the traffic system's safety condition. Conventional 1D TTC may quantify the temporal imminence of a longitudinal collision, while PET characterizes temporal separation within a crossing conflict zone. Neither measure independently captures the number of exposed road users, the potential severity of a collision, the persistence of the hazardous condition, or the operational context in which it occurs. Accordingly, this review organizes the literature using a \emph{risk state}: a continuously updated representation of the current and near-term safety condition of the monitored transportation system, derived from filtered trajectories, scenario-matched SSMs, and contextual information about the conflict environment \citep{mahmud2017application, westhofen2023criticality}.

A risk-state vector at time~$t$ can be written schematically as
\begin{equation}
\mathbf{r}_t = \left[\text{exposure},\ \text{imminence},\ \text{potential severity},\ \text{persistence}\right]^{\top},
\label{eq:risk-state}
\end{equation}
where exposure reflects how many relevant road users or conflict opportunities are present; \emph{imminence} reflects how close an ongoing interaction is to a collision condition, measured with anticipatory, scenario-matched metrics such as TTC, MTTC, PSD, gap time, predicted PET \cite{laureshyn2010evaluation}, or predicted footprint overlap \cite{li20241beyond}; \emph{potential severity} reflects speeds, relative kinetic conditions, road-user type, and whether a vulnerable road user (VRU) is involved; and \emph{persistence} reflects whether the risk is momentary, sustained, or worsening over successive observations. Observed PET and other event-completion measures do not enter imminence, because their values are known only after the event has resolved (see the Temporal Availability subsection below); instead, they contribute to exposure and persistence, for example through the number of low-PET events in the current signal cycle or rolling window. Component definitions and weightings are application-specific; Equation~\eqref{eq:risk-state} is an organizing schema rather than a single mandated implementation. Extreme-value crash-risk estimates offer one rigorous way to populate this representation: a short-term crash probability derived from {2D TTC} extremes summarizes imminence and, through its covariates, part of potential severity over a time window \citep{anis2025realtime, anis2026learning}. The risk state does not replace such estimates. It keeps the four components separate for two reasons. Controllers act on them differently: imminence calls for in-event warnings or evasive constraints, persistence for next-cycle signal changes, and VRU severity for priority rules. And the components draw on inputs that become available at different times, as the Temporal Availability subsection explains.

The risk state functions as the interface between infrastructure perception and downstream decision-making. Rather than requiring a controller to interpret raw point clouds, fragmented trajectories, or a collection of unrelated surrogate measures, the sensing pipeline can supply a structured representation of safety-relevant conditions. Depending on the application, this representation may include conflict exposure, collision imminence, potential severity, and the persistence or rate of change of risk. A downstream controller can then translate the risk state into application-specific objectives or constraints, including signal-timing adjustments, speed harmonization, ramp-metering actions, traveler warnings, or CAV trajectory modifications. The present review focuses on defining this sensing-to-risk interface rather than prescribing a particular control law.

\subsection{Temporal Availability: From Offline Metrics to Real-Time Use}\label{sec:realtime}

SSMs differ not only in the conflict geometries they suit but in when their values become known. TTC, MTTC, PSD, gap time, 2D TTC, and predicted PET \citep[]{laureshyn2010evaluation} are computed from current positions and velocities under an assumed motion model. They therefore exist while an interaction is still unfolding and can support warnings or evasive constraints within the same event.

Observed PET cannot. It is defined only after the first road user has left the conflict area and the second has entered it, so its value is known only once the event it describes has resolved. Observed PET therefore supports online monitoring, such as a rolling count of low-PET events per signal cycle, and can inform control in subsequent cycles. It cannot warn of the same encroachment.

IAPT falls between these cases. It can be evaluated once the encroaching vehicle clears the conflict area and before the through vehicle arrives, but it depends on a predicted arrival time. Table~\ref{tab:ssm-mapping} records this distinction, and Eq.~\eqref{eq:risk-state} applies it: anticipatory measures populate imminence, while event-completion measures populate persistence and exposure.

\section{6. Risk-State-Informed Safety Control: From Metrics to Autonomous Driving}\label{sec:control}

This section develops the control-theoretic side of the pipeline outlined in Figure~\ref{fig:pipeline}. It traces how filtered trajectory data supports kinematic state estimation, how scenario-matched surrogate measures and traffic context are combined into an actionable risk state, and how that risk state is mapped into control objectives and constraints. MPC and related methods appear as illustrative mechanisms rather than as the sole endpoint of the review.

\subsection{Kinematic State Estimation}\label{sec:stateest}

The \customref{Multi-Object Tracking and Trajectory Quality subsection}{sec:tracking} noted that Kalman filtering appears as a near-universal component across roadside LiDAR tracking algorithms, deferring a fuller discussion of why. The reason is that any control-theoretic use of trajectory data depends on a foundation that observational analysis can tolerate more loosely: stable, continuous state estimates. Raw roadside LiDAR detections are inherently susceptible to measurement noise, trajectory fluctuations, and brief occlusions, all of which can distort the SSM calculations introduced in the \customref{Surrogate Safety Measures section}{sec:ssm} badly enough to misrepresent a real conflict as noise, or the reverse. \citet{wen2023research} demonstrate that applying a Kalman filter mitigates positioning deviations and trajectory drift caused by sensor misalignment, enabling adaptive angle calibration of point clouds without manual reference markers. \citet{guan2023evaluation} describe a parallel role for Kalman filtering within tracking itself: predicting object locations in subsequent frames so that target association survives brief line-of-sight loss. Beyond Kalman filtering, the low-pass Savitzky-Golay filtering introduced in the \customref{Multi-Object Tracking and Trajectory Quality subsection}{sec:tracking} through \citet{igene2024integrating} serves the same underlying purpose from a different angle, removing false peaks in velocity and acceleration profiles so that downstream evasive-maneuver or near-miss computations reflect a vehicle's true kinematics rather than sensor artifacts. {For online control, filtering and prediction must be causal and meet the available latency budget; smoothing that uses future samples introduces a delay that must be accounted for. Perception, communication, and computation delays should therefore be reflected in state prediction and safety margins, since smooth trajectories alone do not establish suitability for feedback control.}

\subsection{Risk-State Construction}\label{sec:riskconstruct}

Once kinematic states are stabilized, scenario-matched SSMs and contextual descriptors are combined into the risk-state representation introduced in the \customref{From Surrogate Measures to an Actionable Risk State subsection}{sec:riskstate}. \citet{singh2024conflict} illustrate imminence-focused construction: applying k-means clustering to TTC values extracted from conflict zones, they establish a threshold below {1.15~s---lower than the 1.5~s threshold in Table~\ref{tab:ssm-mapping}, underscoring that thresholds are site- and study-specific---} as signifying a high-risk interaction requiring immediate evasive action. \citet{igene2024integrating} integrate MTTC, the Stopping Distance Index, and the Crash Severity Index into a Rear-End Conflict Index---a composite signal that jointly encodes imminence and severity and predicts crash frequency through {negative binomial regression}. \citet{westhofen2023criticality} review criticality metrics for automated driving more broadly, emphasizing that no single scalar captures every dimension relevant to intervention. Together, these studies show how individual SSM outputs are thresholded, weighted, or fused into structured risk states rather than passed directly to controllers as isolated numbers.

\subsection{Translating Risk States into Control Objectives and Constraints}\label{sec:riskplanning}

Turning a risk state into a control action requires converting exposure, imminence, severity, and persistence into spatial and mathematical boundaries a planning or operations algorithm can respect. \citet{xia2024survey} detail several modeling techniques for this translation, including reachable sets, bounding boxes, driving corridors, and composite risk fields such as the Probabilistic Driving Risk Field and the Geometric Driver Risk Field. {These representations can inform control cost terms or safety constraints, but the mapping from an SSM or risk-state component depends on the vehicle dynamics, prediction model, and treatment of uncertainty. Satisfying a model-based collision-avoidance constraint does not, by itself, establish safety under real-world estimation errors, disturbances, or infeasible control demands.}

\subsection{Illustrative Control Mechanisms}\label{sec:mpc}

With risk expressed as a constraint, MPC is one widely studied mechanism for acting on it: MPC regulates vehicle kinematics over a rolling prediction horizon rather than reacting to a single instantaneous measurement. Other families of controllers---including rule-based signal logic, feedback speed harmonization, and formal safety filters---can consume the same risk-state interface; MPC is emphasized here only because the intersection-coordination literature provides concrete examples of constraint enforcement over a horizon. {The four-component risk state proposed in this review is a conceptual interface for organizing control-relevant information; the MPC studies below illustrate relevant control mechanisms rather than directly validating this interface.}

\citet{mihaly2021performance} demonstrate a centralized MPC algorithm that coordinates multiple autonomous vehicles at non-signalized intersections: by collecting real-time velocity, intention, and position data, the controller calculates the time windows each vehicle will occupy the intersection's conflict zone and continuously applies acceleration commands to prevent those windows from overlapping, minimizing total travel time while strictly enforcing collision-avoidance constraints. \citet{taherian2021autonomous} apply MPC to a related problem with a {linear quadratic regulator} weight transformation, formulated as a discrete-time linear quadratic tracking problem. Their framework introduces Forward and Rear Collision Avoidance Constraints---linear affine constraints that decouple the complex orientation of curved road geometries, allowing a vehicle to execute high-speed evasive maneuvers without destabilizing the surrounding traffic stream or triggering secondary rear-end collisions.

{Control barrier functions (CBFs) provide a complementary mechanism for safety-critical control. \citet{ames2017control} formulate safety through the forward invariance of a prescribed safe set and impose CBF conditions as constraints in a quadratic program. Such a formulation can act as a safety filter that minimally modifies a nominal control command, potentially supplied by an MPC controller. In the infrastructure setting considered here, roadside state estimates and risk information could help inform safety margins or the selection of relevant constraints; however, an SSM or the four-component risk state does not automatically define a valid CBF. The safe set and barrier condition must be constructed using the system dynamics, and any invariance guarantee requires an initially safe state, feasible control inputs, and appropriate treatment of estimation errors and disturbances.} {Related work on force-controlled nonholonomic mobile robots illustrates why barrier construction must account for the underlying dynamics. \citet{han2024safety} combine control Lyapunov functions (CLFs) with CBFs constructed through integrator backstepping in a quadratic-programming framework for safety-critical stabilization. \citet{wang2026further} extend this approach using a global strict Lyapunov function in polar coordinates and a CLF--CBF optimization framework, with simulation and experimental validation. For position-level obstacle-avoidance constraints of relative degree two, \citet{han2026scaling} develop scaling-based reciprocal CBFs that preserve the interior of the original physical safe set while recovering first-order control authority. These studies offer examples of dynamics-aware safety-constraint construction that could inform downstream vehicle control; connecting such constructions to roadside SSM-derived risk information remains an application-specific step.}

\subsection{Application Domains}\label{sec:applications}

The same risk-state interface supports several deployment contexts, differing mainly in which actuators receive the downstream command. \emph{Intersection signal control} can use sustained high exposure or worsening persistence to extend all-red clearance, adjust permissive phasing, or trigger bus-priority overrides. \emph{CAV coordination} can consume infrastructure-derived imminence and severity to complement onboard perception, particularly when line-of-sight is occluded at urban intersections. \emph{Freeway speed harmonization} and \emph{ramp metering} map rear-end imminence along mainline segments into advisory or mandatory speed reductions and merge gaps. \emph{VRU protection} can draw on anticipatory measures such as predicted PET or 2D TTC for in-event warnings, while distributions of observed PET across signal cycles can inform leading pedestrian intervals or protected-turn phasing; in both cases, severity terms that explicitly encode pedestrian or cyclist presence apply. In each case, the review stops at the sensing-to-risk-to-objective chain; selecting and validating a specific control law for a given agency or original equipment manufacturer stack remains application engineering beyond this paper's scope.

Taken together, the \customref{Kinematic State Estimation}{sec:stateest} through \customref{Application Domains}{sec:applications} subsections trace a single path aligned with Figure~\ref{fig:pipeline}: filtered trajectories become stable kinematic estimates, estimates and scenario-matched SSMs become a risk state, the risk state becomes objectives and constraints, and constraints inform actuation across infrastructure and vehicle domains. Where onboard sensing is occluded, degraded, or simply absent from a connected vehicle, infrastructure-derived risk states computed as described here could plausibly supplement or cross-validate a vehicle's own assessment, extending proactive safety from an agency dashboard into the control loop itself.

\section{7. Future Directions}\label{sec:future}

The most direct extension of this paper's argument moves beyond the benchmark-and-dataset framing common to prior roadside-sensing surveys toward the trajectory this paper has argued for directly. Rather than treating roadside deployments as static observation platforms, the field can build closed-loop cycles in which deployment data discovers failure modes, triggers targeted review, refreshes labels, and improves detection and tracking models iteratively: sensing produces trajectories, review identifies hard cases, curated labels improve models, and updated systems return to the field with better rare-event coverage.

The \customref{Risk-State-Informed Safety Control section}{sec:control} argued that this loop need not terminate at an agency dashboard. If infrastructure-derived risk states can be filtered, updated over time, and expressed as spatial or operational constraints reliably enough for field deployment, the same signals are, in principle, a candidate input for connected or automated vehicles operating in range of that infrastructure, particularly when onboard sensing is occluded, degraded, or simply unavailable. Realizing this possibility will require substantial further work on trajectory reliability, uncertainty quantification, and deployment robustness before infrastructure-derived signals could be trusted as an input to a safety-critical control loop; it is a direction this paper argues for rather than one it has demonstrated. But the pipeline traced across the \customref{Sensing Technologies}{sec:sensing} through \customref{Risk-State-Informed Safety Control}{sec:control} sections---from sensor choice, through detection and tracking, through scenario-matched SSMs, to filtered risk states and control objectives---is the shape that demonstration would need to take.

\section{{8. Conclusion}}\label{sec:conclusion}

This paper set out to treat roadside sensing, surrogate safety measurement, and control-theoretic intervention as a single pipeline rather than three literatures that have only recently begun to converge. The \customref{Sensing Technologies section}{sec:sensing} compared camera, LiDAR, and UAV-based sensing specifically for their suitability to safety-relevant trajectory extraction, showing that no sensor dominates every dimension and that sensor choice is itself a safety-relevant design decision rather than a preliminary engineering one. The \customref{Detection, Tracking, and Trajectory Computation section}{sec:detection} showed why that choice matters downstream: detection and tracking pipelines determine whether a deployment produces a stream of actor positions or trajectories stable enough to support conflict inference, with filtering emerging as a near-universal requirement across both LiDAR- and camera-based approaches. The \customref{Surrogate Safety Measures section}{sec:ssm} organized SSMs by conflict geometry and temporal availability, positioned EVT-based crash-risk estimation within that framework, and introduced an actionable risk-state representation that composes exposure, imminence, potential severity, and persistence from filtered, scenario-matched metrics. The \customref{Risk-State-Informed Safety Control section}{sec:control} extended this framework by tracing how risk states are translated into control objectives and constraints and how illustrative mechanisms such as MPC and several infrastructure and CAV application domains consume that interface, without treating any single controller as the review's endpoint.

The throughline connecting all four sections is that validity compounds rather than resets at each stage. A well-chosen SSM computed from unstable trajectories is not a safety measure at all, and a well-tuned MPC controller acting on an uncalibrated risk signal inherits every upstream weakness in the pipeline that produced it. This is the sense in which sensing, tracking, metrics, and control are inseparable rather than sequential add-ons to one another.

Two implications follow for the audiences this paper has kept in view throughout. For transportation agencies, the practical takeaway is that sensing and metric selection should be matched to the specific conflict types a site presents, rather than defaulting to a single sensor or a single threshold across an entire deployment. For the autonomous vehicle research community, the more speculative but consequential implication is that infrastructure-derived risk signals represent an underexplored complement to onboard sensing, particularly at the occluded or degraded-visibility intersections where onboard perception alone is least reliable, provided the reliability and uncertainty concerns raised in the \customref{Future Directions section}{sec:future} are addressed first. Realizing that possibility remains future work, but the pipeline this paper has traced, from a roadside point cloud to a steering constraint, is the foundation it would need to be built on.

\newpage
\bibliographystyle{chicago}
\bibliography{references}

\end{document}